\documentclass{article}%
\usepackage{amsmath}
\usepackage{amsfonts}
\usepackage{amssymb}
\usepackage{graphicx}
\usepackage{natbib}
\usepackage{xcolor,xspace,colortbl,ragged2e,rotating}
\usepackage[colorlinks,linkcolor=blue,citecolor=blue,urlcolor=red,bookmarks=false,hypertexnames=true]%
{hyperref}%
\providecommand{\U}[1]{\protect\rule{.1in}{.1in}}

\newtheorem{assumption}{Assumption}

\newtheorem{corollary}{Corollary}

\newtheorem{proposition}{Proposition}

\newtheorem{remark}{Remark}

\ifx\pdfoutput\relax\let\pdfoutput=\undefined\fi
\newcount\msipdfoutput
\ifx\pdfoutput\undefined\else
\ifcase\pdfoutput\else
\ifx\paperwidth\undefined\else
\ifdim\paperheight=0pt\relax\else\pdfpageheight\paperheight\fi
\ifdim\paperwidth=0pt\relax\else\pdfpagewidth\paperwidth\fi
\fi\fi\fi
\begin{document}

\title{VAR models with an index structure: A survey with new results}
\author{Gianluca Cubadda\thanks{ Universita' di Roma "Tor Vergata", Dipartimento di
Economia e Finanza, Via Columbia 2, 00133 Roma, Italy. Email:
gianluca.cubadda@uniroma2.it. }\\Universit{\`a} di Roma "Tor Vergata"}
\date{
\today
\footnote{ Previous versions of this paper were presented at the Intermediate
workshop on methodological and computational issues in large-scale time series
models for economics and finance in Messina, the The Villa Mondragone time
series symposium in honour of Marco Lippi in Monte Porzio Catone (Rome), and
the 11$^{\text{th}}$ ICEEE in Palermo. The author thanks the participants for
their helpful comments and suggestions. The usual disclaimers apply. The
financial support of MUR under the 20223725WE (PRIN 2022) grant is gratefully
acknowledged.}}
\maketitle

\begin{abstract}
The main aim of this paper is to review recent advances in the multivariate
autoregressive index model [MAI], originally proposed by
\cite{reinsel1983some}, and their applications to economic and financial time
series. MAI has recently gained momentum because it can be seen as a link
between two popular but distinct multivariate time series approaches: vector
autoregressive modeling [VAR] and the dynamic factor model [DFM]. Indeed, on
the one hand, the MAI is a VAR model with a peculiar reduced-rank structure;
on the other hand, it allows for identification of common components and
common shocks in a similar way as the DFM. The focus is on recent developments
of the MAI, which include extending the original model with individual
autoregressive structures, stochastic volatility, time-varying parameters,
high-dimensionality, and cointegration. In addition, new insights on previous
contributions and a novel model are also provided.

\bigskip


\end{abstract}

\newpage


\section{Introduction}

The Vector Auto-Regressive model [VAR] and the Dynamic Factor Model [DFM] are
arguably among the most popular tools for analyzing economic and financial
variables over time. Since the seminal contribution of
\cite{sims1980macroeconomics}, VARs have been theoretically extended and
practically implemented to forecast, perform structural analysis, and detect
comovements in multivariate time series. DFMs were introduced more recently
(\cite{forni2000generalized}, \cite{Forni_Lippi2001},
\cite{stock2002forecasting}, \cite{stock2002macroeconomic}, \cite{bai_ng2003},
and \cite{bai2003}), but rapidly contested the role of the workhorse in
empirical macroeconomics.

The main reason for the success of the DFM is two-fold. First, it allows
handling a much larger number of variables than those that are generally
employed in traditional small-scale VARs, thus potentially boosting
forecasting accuracy and solving the informational deficiency problems that
arise in structural analyses when the agent's information set is richer than
the econometrician's information set (see, e.g., \cite{Forni2014}). Second,
DFM allows one to disentagle the shocks that drive the common components of
several time series and recover the structural shocks from those common
shocks. Hence, in structural DFMs, the number of shocks is smaller than the
number of variables (\cite{ForniET}), which is in line with dynamic stochastic
general equilibrium models (see \cite{DSGE2016} and the references therein)
and, more generally, with the standard macroeconomic view that a small number
of shocks drives aggregate fluctuations.

Efforts have recently been made to endow the VAR with the above-mentioned
features of the DFM. On the one hand, shrinkage estimators have been proposed
for medium-large VARs, both from a Bayesian perspective (e.g.,
\cite{banbura2010large}, \cite{koop2013forecasting}, and
\cite{carriero2015bayesian}) and from a classical standpoint (e.g.,
\cite{Hsu2008}, \cite{KC2005}, and \cite{HMSJoFE}). On the other hand, the
Multivariate Autoregressive Index (MAI) model -- originally proposed by
\cite{reinsel1983some} as a convenient approach to dimension reduction in
stationary VARs -- has recently gained renewed attention.\footnote{At the end
of 2024, the annual citation rate of \cite{reinsel1983some} in Scopus has
increased by about 54\% in the last 9 years, with the majority of recent
citations coming from econometric journals.}.

Late advances have shown that MAI and its variants allow for both forecasting
variables and identifying shocks analogously to the DFM but without
encountering some issues in model identification and statistical inference
that characterize the latter, such as the requirement that the number of
variables diverges at a given rate and the need of specific assumptions on
both the correlation structure of the idiosyncratic components and the factor
loadings (see, \textit{i.a.}, \cite{bai2003}, and \cite{bai2006confidence}).
Moreover, VARs with index structures have been shown to be able to accommodate
features such as stochastic volatility (\cite{carriero2022global}) and
time-varying parameters (\cite{CGG2025}), which are not easy to handle within
the DFM framework.

The MAI falls within reduced-rank VARs, a general class of models that include
as special cases both the cointegrated VAR (see \cite{johansen1995likelihood}
and the references therin) and the Common Serial Correlation [CSC] models (see
\cite{CH2022ORE} and the references therin). Although CSC models and MAI have
similar mathematical formulations, their respective goals and properties are
rather different; whereas the former are based on the existence of (possibly
dynamic) linear combinations of autocorrelated time series that are white
noise (\cite{EK1993}, \cite{Vahid1993}, \cite{Cubadda2007},
\cite{carriero2011forecasting}, \cite{cubadda2011}, \cite{Bernardini2015}),
VARs with an index structure assume that there is a limited number of channels
through which information from the past is transmitted to the variables of interest.

The main aim of this paper is twofold. First, recent developments in MAI are
reviewed, such as the structural MAI (\cite{carriero2016structural}), the
vector heterogeneous index model for realized volatilities
(\cite{cubadda2017vector}), as well as augmentations of the original model
with individual autoregressive structures (\cite{cubadda2019representation}),
stochastic volatility (\cite{carriero2022global}), time-varying parameters
(\cite{CGG2025}), high-dimensionality (\cite{cubadda2022}), and cointegration
(\cite{CM2024}). Second, new results are provided in terms of representation
theory for the various models, and a novel modeling is proposed, namely the
cointegrated index-augmented autoregressive model, which combines and extends
the results in \cite{cubadda2019representation} and \cite{CM2024}.

This paper is organized as follows. Focusing on representation theory, Section
2 reviews previous contributions and provides new insights into some of them.
Section 3 presents the new model and deals with its estimation, whereas some
details of the estimation procedure are relegated to the Appendix. Finally,
Section 4 provides some conclusions.


\section{VAR Models with an Index Structure}

In this Section we review the models that are rooted from the original MAI
formulation and provide new results on representation theory of some of them.
Analogies and differences with the DFM are discussed in detail. Estimation and
identification issues are also covered.

\subsection{The Stuctural Multivariate Autoregressive Index Model}

Let us assume that the $n-$vector time series $Y_{t}=(y_{1t},...,y_{nt}%
)^{^{\prime}}$ is generated by the following stationary VAR$(p)$ model:
\begin{equation}
\Phi(L)Y_{t}=\varepsilon_{t},\text{ \ \ }t=1...T, \label{VAR}%
\end{equation}
where $L$ is the lag operator, $\Phi(L)=$ $I_{n}-\sum_{j=1}^{p}\Phi_{j}L^{j}$,
and $\varepsilon_{t}$ is a vector or $n$ errors with $\mathrm{E}%
(\varepsilon_{t}\varepsilon_{t}^{\prime})=\Sigma$ (positive definite) and
finite fourth moments, $\mathrm{E}(\varepsilon_{t}|\digamma_{t-1})=0$, and
$\digamma_{t}$ is the natural filtration of the process $Y_{t}$. For
simplicity, deterministic elements are ignored.

The key assumption of MAI (\cite{reinsel1983some}) is the following:

\begin{assumption}
It holds
\[
\lbrack\Phi_{1}^{\prime},\ldots,\Phi_{p}^{\prime}]^{\prime}=[\alpha
_{1}^{\prime},\ldots,\alpha_{p}^{\prime}]^{\prime}\omega^{\prime},
\]
where $\omega$ is a full-rank $n\times q-$matrix with $q<n$, and $\alpha_{j}$
is a $n\times q-$matrix for $j=1,\ldots,p$.
\end{assumption}

Under Assumption 1, Model (\ref{VAR}) can be rewritten as
\begin{equation}
Y_{t}=\sum_{j=1}^{p}\alpha_{j}\underset{f_{t-i}}{\underbrace{\omega^{\prime
}Y_{t-i}}}+\varepsilon_{t} \label{MAI}%
\end{equation}
where linear combinations $f_{t}=\omega^{\prime}Y_{t}$ are called the indexes.
The MAI has at most $nq(p+1)-q^{2}$ mean parameters, which implies a
significant dimension reduction when $p$ is small w.r.t. $n$.\footnote{Indeed,
the matrix $\omega$, once identified through normalizing restrictions, has
$q(n-q)$ free parameters.}

By premultiplying with $\omega^{\prime}$ both sides of Equation (\ref{MAI}) we
get
\begin{equation}
f_{t}=\sum\limits_{j=1}^{p}\omega^{\prime}\alpha_{j}f_{t-j}+\omega^{\prime
}\varepsilon_{t}, \label{Indexes}%
\end{equation}
which shows that the indexes follow a VAR$(p)$ process and not a VARMA, as is
generally the case for linear combinations of elements of a VAR (see
\cite{Cubadda2009} and the references therein).

\begin{remark}
In view of Equations (\ref{MAI}) and (\ref{Indexes}), the MAI resembles the
exact DFM [EDFM] (see \cite{Lippi2ORE} and the references therein), but there
are also some relevant differences. First, in the EDFM series $Y_{t}$ load the
factors even contemporaneously and not only with lags. Second, the factors and
the idiosyncratic terms in the EDFM are uncorrelated at any lag-lead, whereas
in the MAI we have $\mathrm{E}(f_{t}\varepsilon_{t+j})=0$ only for $j>0$.
Third, the contemporaneous variance matrix of the idiosyncratic terms in the
EDFM is diagonal, whereas $\Sigma$ is generally not.
\end{remark}

Putting emphasis on the analogies between MAI and EDFM,
\cite{carriero2016structural} propose identifying structural shocks as linear
transformations of the index shocks only. Starting from the Wold
representation of series $Y_{t}$
\[
Y_{t}=\Psi(L)\varepsilon_{t},
\]
and inserting between $\Psi(L)$ and $\varepsilon_{t}$ the decomposition of the
identity matrix as in \cite{centoni2003measuring}
\begin{equation}
I_{n}=\Sigma\omega(\omega^{\prime}\Sigma\omega)^{-1}\omega^{\prime}%
+\omega_{\perp}(\omega_{\perp}^{\prime}\Sigma^{-1}\omega_{\perp})^{-1}%
\omega_{\perp}^{\prime}\Sigma^{-1}, \label{CC}%
\end{equation}
one gets the following decomposition of series $Y_{t}:$%
\begin{equation}
Y_{t}=\chi_{t}+\iota_{t} \label{decomp}%
\end{equation}
where
\begin{align}
\chi_{t}  &  =\Psi(L)\Sigma\omega\underline{\Sigma}^{-1}\varepsilon_{t}^{\chi
},\label{Chi}\\
\iota_{t}  &  =\Psi(L)\omega_{\perp}(\omega_{\perp}^{\prime}\Sigma^{-1}%
\omega_{\perp})^{-1}\varepsilon_{t}^{\iota}, \label{Iota}%
\end{align}
$\underline{\Sigma}=\omega^{\prime}\Sigma\omega$, $\varepsilon_{t}^{\chi
}=\omega^{\prime}\varepsilon_{t}$, $\varepsilon_{t}^{\iota}=\omega_{\perp
}^{\prime}\Sigma^{-1}\varepsilon_{t}$, $\mathrm{E}(\varepsilon_{t}^{\chi
}\varepsilon_{t}^{\iota\prime})=0$, and $\mathrm{E}(\chi_{t}\iota
_{t-j}^{\prime})=0$ for $\forall j$.

Since the shocks $\varepsilon_{t}^{\chi}$ are those of the indexes,
$\varepsilon_{t}^{\chi}$ may be interpreted as the common shocks and $\chi
_{t}$ as the common components of the series $Y_{t}$. Similarly,
$\varepsilon_{t}^{\iota}$ and $\iota_{t}$ can be labeled, respectively, as
uncommon shocks and a uncommon component.

Interestingly, post-multiplying with $\omega_{\perp}$\ both sides of the
relation%
\[
\Psi(L)(I_{n}-\sum_{j=1}^{p-1}\alpha_{j}\omega^{\prime}L^{j})=I_{n}%
\]
we get $\Psi(L)\omega_{\perp}=\omega_{\perp}$, which in turn implies that the
Wold polynomial matrix of the MAI has the form%
\begin{equation}
\Psi(L)=I_{n}+\sum_{j=1}^{\infty}\theta_{j}\omega^{\prime}L^{j} \label{Psi(L)}%
\end{equation}
where $\theta_{j}$ is an $n\times q-$matrix for $j>0$.

Having substituted $\Psi(L)$ in Equations (\ref{Chi}) and (\ref{Iota}) with
the RHS of Equation (\ref{Psi(L)}), we can finally prove the following proposition:

\begin{proposition}
In the MAI, the components of $Y_{t}$\ in (\ref{decomp}) read%
\begin{align*}
\chi_{t}  &  =(\Sigma\omega\underline{\Sigma}^{-1}+\sum_{j=1}^{\infty}%
\theta_{j}L^{j})\varepsilon_{t}^{\chi},\\
\iota_{t}  &  =\omega_{\perp}(\omega_{\perp}^{\prime}\Sigma^{-1}\omega_{\perp
})^{-1}\varepsilon_{t}^{\iota},
\end{align*}
where the uncommon component $\iota_{t}$ is a $n-$dimensional white noise such
that $\mathrm{Rank}\left(  \mathrm{E}(\iota_{t}\iota_{t}^{\prime})\right)
=n-q$.
\end{proposition}

\begin{corollary}
The indexes and the common component are linked through the relation
$f_{t}=\omega^{\prime}\chi_{t}$, which trivially follows from Proposition 1.

\begin{remark}
In view of Proposition 1, the decomposition (\ref{decomp}) has clear analogies
with the analogous decomposition in the EDFM. However, differently from the
idiosyncratic terms in the EDFM, the uncommon component $\iota_{t}$ is
obviously cross-sectionally dependent.\footnote{Remarkably, when the factors
in the EDFM are estimated by some principal components of series $Y_{t}$, the
sample variance matrix of the estimated idiosyncratic component has
reduced-rank as well.}
\end{remark}
\end{corollary}

\cite{carriero2016structural} suggest to recover the structural shocks as
linear transformations of the common shocks $\varepsilon_{t}^{\chi}$ only.
Hence, at most $q<n$ structural shocks can be recovered, as occurs in DFMs and
in dynamic stochastic general equilibrium models. In principle, all the
identification strategies that are available for structural VARs or structural
DFMs (see \cite{SW2016chapter} and the references therein) can be adopted.

On the estimation side, \cite{carriero2016structural} prove that the iterative
maximum likelihood procedure proposed by \cite{reinsel1983some} is consistent
when $n=o(\sqrt{T})$. Moreover, they provide an MCMC algorithm for Bayesian
estimation and show by simulations that the Bayesian approach outperforms the
classical one when $n=15,20$. Finally, they document the practical value of
the structural MAI by two empirical applications, on the transmission
mechanism of monetary policy and on the propagation of demand and supply shocks.

\subsection{{The Vector Heterogeneous Autoregressive Index Model}}

The univariate Heterogeneous AR model [HAR], originally proposed by
\cite{Corsi2009}, is a popular tool to analyze and forecast daily realized
volatility [RV] measures without resorting to more involved long-memory
models. Technically speaking, the HAR is a constrained AR$(22)$ model where
the predictors are the first lags of: (i) the daily RV; (ii) the weekly (5
days) average of the daily RV; (iii) the monthly (22 days) average of the
daily RV.

\cite{cubadda2017vector} propose a multivariate HAR for a set of $n$ daily RV
measures $Y_{t}^{(d)}\equiv\left(  Y_{1,t}^{(d)},\ldots,Y_{n,t}^{(d)}\right)
^{\prime}$ that is endowed with an index structure. In particular, the Vector
Heterogeneous Autoregressive Index model [VHARI] reads as follows
\[
Y_{t}^{(d)}=\alpha^{(d)}\omega^{\prime}Y_{t-1d}^{(d)}+\alpha^{(w)}%
\omega^{\prime}Y_{t-1d}^{(w)}+\alpha^{(m)}\omega^{\prime}Y_{t-1d}%
^{(m)}+\varepsilon_{t},
\]
where $(d)$, $(w)$, and $(m)$ denote, respectively, time horizons of one day,
one week, and one month such that%
\[
Y_{t}^{(w)}=\frac{1}{5}\sum_{j=0}^{4}Y_{t-jd}^{(d)},\text{ \ \ }Y_{t}%
^{(m)}=\frac{1}{22}\sum_{j=0}^{21}Y_{t-jd}^{(d)}%
\]

The VHARI enjoys two important properties that are not shared by alternative
approaches to inducing dimension reduction in the Vector HAR\footnote{The most
obvious alternatives to the VHARI likely are multivariate principal component
regression and reduced-rank regression.}: First, the indexes $f_{t}%
^{(d)}=\omega^{\prime}Y_{t-1d}^{(d)}$ preserve the temporal cascade structure
of the HAR model since
\[
f_{t}^{(w)}=\frac{1}{5}\sum_{j=0}^{4}f_{t-jd}^{(d)},\text{ \ \ }f_{t}%
^{(m)}=\frac{1}{22}\sum_{j=0}^{21}f_{t-jd}^{(d)},
\]
Second, pre-multiplying both sides of the VHARI by $\omega^{\prime}$ yields
the following:
\[
f_{t}^{(d)}=\omega^{\prime}\alpha+\omega^{\prime}\alpha^{(d)}f_{t-1d}%
^{(d)}+\omega^{\prime}\alpha^{(w)}f_{t-1d}^{(w)}+\omega^{\prime}\alpha
^{(m)}f_{t-1d}^{(m)}+\omega^{\prime}\varepsilon_{t},
\]
which shows that the indexes follow a multivariate HAR model. In particular,
when $q=1$ a univariate HAR model generated all the dynamics of the $n$ RVs.

On the estimation side, \cite{cubadda2017vector} suggest using a switching
algorithm [SA], an iterative method for numerical maximization of the
log-likelihood of complex models that has a long tradition in time-series
analysis (see \cite{Boswijk2004} and the references therein). In particular,
the proposed SA requires the following steps:

\begin{enumerate}
\item Given an (initial) estimate of $\omega$, maximize the conditional
Gaussian likelihood $\ell(A,\Sigma|\omega)$ where $A=[\alpha^{(d)^{\prime}%
},\alpha^{(w)^{\prime}},\alpha^{(m)^{\prime}}]^{\prime}$.

\item Given the previously obtained estimates of $A$ and $\Sigma$, maximize
the conditional likelihood $\ell(\omega|A,\Sigma)$.

\item Repeat Steps 1 and 2 until numerical convergence occurs.\footnote{A
general proof of the convergence of this family of iterative procedures is
given by \cite{Oberhofer1974}.}
\end{enumerate}

A key point of the above SA is that both Steps 1 and 2 require running OLS
regressions only. This feature provides the SA with several advantages over
Newton-type optimization methods, such as computational simplicity, no need
for normalization conditions in $\omega$, explicit optimization at each step,
and ease of application of regularization schemes or linear restrictions on
parameters (see \cite{cubadda2019representation} for additional discussion).
Furthermore, when the SA is initialized with consistent estimates and is
iterated sufficiently often, the resulting estimator is asymptotically
equivalent to the ML one (\cite{HautschJoFE}). \cite{cubadda2017vector} show
by simulation that the suggested SA performs well even when elements of
$\varepsilon_{t}$ have a log-normal error distribution with GARCH variances.

Following \cite{Patton2009}, \cite{cubadda2017vector} use the VHARI to build
the optimal linear combination of ten different estimators of the volatility
of the same market and to evaluate its merits through an out-of-sample
forecasting exercise. The VHARI model proves to work well, often outperforming
previously existing methods.

\subsection{{The Index-Augmented Auto-Regressive Model}}

A possible limitation of MAI as a forecasting tool is that the only predictors
of the series $y_{i,t}$, for $i=1,\ldots,n$, are the lagged indexes, whereas
the forecasts obtained through the DFM exploit information coming from the
past of both factors and the series $y_{i,t}$ itself (see the seminal
contributions by \cite{stock2002forecasting} and \cite{stock2002macroeconomic}%
). Although the indexes may be interpreted as 'supervised' factors that are
constructed for emphasizing the comovements between the present and the past
of the system, it may occur that some variables are better predicted by their
own lags rather than by any linear combination of all variables only.

In order to overcome such limitation, \cite{cubadda2019representation}
extended the basic MAI model by allowing individual AR structures for each
element of $Y_{t}$. Their key assumption is the following.

\begin{assumption}
It holds%
\[
\phi_{ik}^{(j)}=%
{\textstyle\sum\limits_{m=1}^{q}}
\alpha_{im}^{(j)}\omega_{km},
\]
where $\phi_{ik}^{(j)}$ is the generic element of the polynomial matrix
$\Phi_{j}$, $\omega_{km}$ is the generic element of $\omega$, and $\alpha
_{im}^{(j)}$ is the generic element of $\alpha_{j}$ for $j=1,\ldots,p$,
$i=1,\ldots,n$, $k=1,\ldots,i-1,i+1,\ldots,n$.
\end{assumption}

In words, Assumption 2 states that there is a reduced number of channels $p$
through which each variable is influenced by the past of other variables in
the system, which is consistent with the common view that few shocks are
responsible for most macroeconomic fluctuations.

Under Assumption 2 and using the reparametrization $\delta_{ii}^{(j)}%
=\phi_{ii}^{(j)}-\sum_{m=1}^{q}\alpha_{im}^{(j)}\omega_{im}$, Model
(\ref{VAR}) can be rewritten into the following {Index-Augmented
Auto-Regressive model} [IAAR]:%
\begin{equation}
Y_{t}=\sum_{j=1}^{p}D_{j}Y_{t-j}+\sum\limits_{j=1}^{s}\alpha_{j}%
f_{t-j}+\varepsilon_{t}, \label{MIAAR}%
\end{equation}
where $D_{j}$\ is a $n\times n$ diagonal matrix with $\delta_{ii}^{(j)}$ as
generic diagonal element, and, for greater generality, $s\leq p$.

\begin{remark}
Since the number of parameters of Model (\ref{MIAAR}) is equal to
$n(qs+q+p)-q^{2}$, it is necessary to impose proper upper bounds to either $q$
or $s$ to ensure that the MIAAR is more parsimonious than the VAR. To this
end, it is easy to see that sufficient conditions are $q<n-1$ for $s=p\geq2$
or $s<p-1$ for any $p$ and $q<n$. However, in empirical applications, the
estimated values of $q$ are typically much smaller than $n$ (see
\cite{cubadda2019representation} and \cite{carriero2022global}).
\end{remark}

\begin{remark}
The individual forecasting equation of the IAAR reads%
\begin{equation}
y_{it+1}=\sum_{j=0}^{p-1}\delta_{ii}^{(j)}y_{it-j}+\sum\limits_{j=0}%
^{s-1}\alpha_{i\cdot}^{(j)\prime}f_{t-j}+\varepsilon_{it+1}, \label{IAAR}%
\end{equation}
where $\alpha_{i\cdot}^{(j)\prime}$ is the $i-$th row of matrix $\alpha_{j}$.
Equation (\ref{IAAR}) is entirely analogous to the individual forecasting
equation of the DFM, with one important difference. Whereas factors are
typically estimated using principal component methods, which aim to maximize
the contemporaneous variability of series $Y_{t}$, the indexes in (\ref{IAAR})
are constructed explicitly taking into account the covariability between each
series $y_{it}$ and the lags of other elements of $Y_{t}$ conditionally on the
lags of the series $y_{it}$.
\end{remark}

\begin{remark}
Interestingly, by the same argument underlying Proposition 1, we see that,
differently from the MAI, $\Psi(L)\omega_{\perp}\neq\omega_{\perp}$, which
implies, in view of Equation (\ref{Iota}), that the uncommon component
$\iota_{t}$\ is generally autocorrelated in the case of the IAAR. Hence, the
decomposition (\ref{decomp}) for the IAAR closely resembles the analogous
decomposition in the approximate DFM (see \cite{Lippi2ORE} and the references
therein). However, estimation of the indexes $f_{t}$ does not require that
$n\rightarrow\infty$ and any condition on the autocorrelations and
cross-correlations of elements of $\iota_{t}$ or on the loadings $\alpha_{j}$
as in the approximate DFM.
\end{remark}

\cite{cubadda2019representation} proposed a two-step SA for the estimation of
the IAAR, along with a variant where a $\ell_{2}$ regularization scheme is
applied in both steps. They show by simulations that the regularized version
of the SA outperforms the standard one with $n=20$. Regarding model
specification, they opt for the use of Information Criteria [IC], in line with
previous contributions showing that IC outperform likelihood ratio tests in
selection of reduced-rank VAR models (see, e.g., \cite{Gonzalo1999chapter},
\cite{CavaliereOBES}, \cite{CavaliereET}). Finally, the IAAR proves to
outperform well-known macroeconomic forecasting methods when applied to
systems with $n$ ranging from $4$ to $40$.

\cite{carriero2022global} endowed the IAAR with Stochastic Volatility
[IAAR-SV] in the errors $\varepsilon_{t}$ and offered Bayesian estimation
using Markov Chain Monte Carlo [MCMC] techniques. Furthermore, they use
(\ref{CC}) to decompose the time-varying volatility $\mathrm{E}(\varepsilon
_{t}\varepsilon_{t}^{\prime})=\Sigma_{t}$\ as follows:%
\[
\Sigma_{t}=\underset{\mathrm{common}}{\underbrace{\Sigma_{t}\omega
(\omega^{\prime}\Sigma_{t}\omega)^{-1}\omega^{\prime}\Sigma_{t}}%
}+\underset{\mathrm{uncommon}}{\underbrace{\omega_{\perp}(\omega_{\perp
}^{\prime}\Sigma_{t}^{-1}\omega_{\perp})^{-1}\omega_{\perp}^{\prime}}}%
\]

\cite{carriero2022global} apply the IAAR-SV to analyze the commonality in both
levels and volatilities of inflation rates in several countries, and their
main finding is that a substantial fraction of inflation volatility can be
attributed to a global factor that also drives inflation levels and their persistence.

\subsection{{The Time-Varying Multivariate Autoregressive Index Model}}

A further step towards taking into account parameter instabilities over time
was taken by \cite{CGG2025}, who proposed the following MAI with Time Varying
Parameters and Time-Varying Volatility [MAI-TVP-TVV]:
\begin{align*}
Y_{t}  &  =\sum_{j=1}^{p}\alpha_{j,t}\omega^{\prime}Y_{t-i}+\varepsilon_{t},\\
\boldsymbol{\alpha}_{t}  &  =\boldsymbol{\alpha}_{t}+\kappa_{t}%
\end{align*}
where $\boldsymbol{\alpha}_{t}=\mathrm{Vec}(\alpha_{1,t}^{\prime}%
,\ldots,\alpha_{n,t}^{\prime})^{\prime}$, $\varepsilon_{t}\sim N(0,\Sigma
_{t})$, $\kappa_{t}\sim N(0,Q_{t})$, $\varepsilon_{t}$ and $\kappa_{t}$ are
independent at any lag and lead. Notice that it is assumed that the index
loadings evolve over time as random walks, while the index weights $\omega$
remain stable.

In order to overcome the computational limitation related to MCMC procedures,
\cite{CGG2025} offers a hybrid estimation method that combines the SA, Kalman
filter with forgetting factors (\cite{KK2014}), and exponentially weighted
moving average techniques (\cite{JOPSB2023}) for the time-varying volatility.

An empirical application, where 25 US quarterly time series are used to
forecast three key macroeconomic variables, shows that the MAI-TVP-TVV is one
of the best models in a large set of competitors for all targets, improving
upon its counterparts especially at short horizons. Other interesting findings
are that, once the MAI is endowed with time-varying volatility [MAI-TVV],
there are no clear improvements in adding time-varying parameters for point
forecasting, but the MAI-TVP-TVV always outperforms the MAI-TVV in density forecasting.

\subsection{{The Dimension-Reducible VAR}}

\cite{cubadda2022} studied the conditions under which the dynamics in a
large-dimensional VAR are entirely generated by a small-scale VAR. They show
that such conditions are met when the coefficient matrices of the large VAR
have the same common right space and a common left null space. This entails
combing Assumption 1 with the following:

\begin{assumption}
It holds%
\[
\omega_{\perp}^{\prime}[\Phi_{1},\ldots,\Phi_{p}]=0
\]

\end{assumption}

Assumption 3 is popularly known in time series econometrics as the CSC (see
\cite{CH2022ORE} and the reference therein) given that
\[
\omega_{\perp}^{\prime}Y_{t}=\omega_{\perp}^{\prime}\varepsilon_{t},
\]
that is, there exist $(n-q)$ linear combinations of variables $Y_{t}$ that are
white noise and, as such, cannot exhibit cyclical behavior.

Taking Assumptions 1 and 3 together leads to the Dimension-Reducible VAR model
[DRVAR]:
\begin{equation}
Y_{t}=\sum_{j=1}^{p}\omega\phi_{j}f_{t-j}+\varepsilon_{t}, \label{DRVAR}%
\end{equation}
where $\phi_{j}$ is a $q\times q$ matrix for $j=1,...,p$.

Assuming, without loss of generality, that $\omega^{\prime}\omega=I_{q}$ and
$\omega_{\bot}^{\prime}\omega_{\bot}=I_{n-q}$, we can decompose series $Y_{t}$
as follows%
\begin{equation}
Y_{t}=\omega f_{t}+\omega_{\bot}\eta_{t}, \label{decomp_3}%
\end{equation}
where $f_{t}$ is the dynamic component and $\eta_{t}=\omega_{\bot}^{\prime
}\varepsilon_{t}$ is the static one. Premultiplying both sides of DRVAR by
$\omega^{\prime}$\ one gets%
\[
f_{t}=\sum_{j=1}^{p}\phi_{j}f_{t-j}+\varepsilon_{t}^{\chi},
\]
where $\varepsilon_{t}^{\chi}=\omega^{\prime}\varepsilon_{t}$, which shows
that $f_{t}$\ is generated by a $q-$dimensional VAR($p$) process.

By inserting the Wold representation of the dynamic components $f_{t}$\ in
Equation (\ref{decomp_3}) it follows:
\begin{equation}
Y_{t}=\omega\gamma(L)\varepsilon_{t}^{\chi}+\omega_{\bot}\eta_{t},
\label{R_Wold}%
\end{equation}
where $\gamma(L)^{-1}=I_{n}-\sum_{j=1}^{p}\phi_{j}L^{j}$. Finally, by linearly
projecting $\omega_{\bot}\eta_{t}$ on $\varepsilon_{t}^{\chi}$, we obtain
$\omega_{\bot}\eta_{t}=\rho\varepsilon_{t}^{\chi}+\nu_{t}$ with $\mathrm{E}%
(\varepsilon_{t}^{\chi}v_{t}^{\prime})=0$, which can be inserted into Equation
(\ref{R_Wold}) to get%
\begin{equation}
Y_{t}=C(L)\varepsilon_{t}^{\chi}+\nu_{t}, \label{Sdecomp}%
\end{equation}
where $C_{0}=\omega+\rho$\ and $C_{j}=\omega\gamma_{j}$ for $j>0$.

Representation (\ref{Sdecomp}) highlights that system dynamics are completely
generated by common reduced form errors $\varepsilon_{t}^{\chi}$.
Consequently, \cite{cubadda2022} label $\nu_{t}$ as the ignorable errors, as
they are noise without structural interpretation. Since the errors
$\varepsilon_{t}^{\chi}$\ and $\nu_{t}$ are uncorrelated at any lead and lag,
it is then possible to recover the structural shocks solely from the reduced
form errors $\varepsilon_{t}^{\chi}$ of the common component $\chi_{t}$ using
any of the procedures that are commonly employed in structural VARs or
structural DFMs (see \cite{SW2016chapter} and the references therein).

In order to estimate the matrix $\omega$, one may rely on a nonparametric
estimator proposed by \cite{Lam2011}. The underlying intuition is that the
matrix $\omega$ lies in the space generated by the eigenvectors associated
with the $q$ nonzero eigenvalues of the symmetric and semipositive definite
matrix.%
\[
M=\sum_{j=1}^{p_{0}}\Sigma_{y}(j)\Sigma_{y}(j)^{\prime},
\]
where $p_{0}\geq p$ and $\Sigma_{y}(j)$ is the autocovariance matrix of series
$Y_{t}$ in lag $j$. Under some regularity conditions, the matrix formed by the
eigenvectors associated with the $q$ largest eigenvalues of the sample
estimate of $M$ is a $\sqrt{T}$-consistent estimator of $\omega$ (up to an
orthonormal transformation) when $q$ is fixed, $n,T\rightarrow\infty$, and
$\omega_{i}^{\prime}\omega_{i}=O(n)$ for $i=1,...,q$, where $\omega
=[\omega_{1},...,\omega_{q}]$. Remarkably, the speed of convergence of the
estimator, namely $\sqrt{T}$, is the same as when the dimension $n$ is finite.

Moreover, \cite{cubadda2022} provide both the OLS and the GLS estimators of
the coefficients $\phi$' s in equation (\ref{DRVAR}) and consistent
information criteria for the selection of $q$, and show by simulations that
the proposed methodology works well with the temporal and cross-sectional
sizes that are typical in macroeconomics. Finally, the approach is applied to
analyze a large set of US economic time series and to identify the shock that
is responsible for most of the common volatility in the business cycle
frequency band.

\subsection{{The Vector-Error Correction Index Model}}

The models considered so far do not explicitly deal with the possible presence
of unit roots. Given that most macroeconomic and financial time series are
characterized by stochastic trends, it is important to understand how a
cointegrated VAR model can be augmented with an index structure.

Let us assume that series $Y_{t}$ follow the Vector Error-Correction Model
[VECM]%
\begin{equation}
\Delta Y_{t}=\alpha_{0}\beta^{\prime}Y_{t-1}+\sum_{j=1}^{p-1}\Pi_{j}\Delta
Y_{t-j}+\varepsilon_{t}, \label{VECM}%
\end{equation}
where $\alpha_{0}$ and $\beta$ are full-rank $n\times r$ ($r<n$) matrices such
that $\alpha_{0}\beta^{\prime}=-\Phi(1)$, $\Pi_{j}=-%
{\textstyle\sum\limits_{i>j}}
\Phi_{i}$ for $j=1,\ldots,p-1$, $\alpha_{0}{}_{\perp}^{\prime}\bar{\Pi}%
\beta_{\perp}$ is non-singular, and $\bar{\Pi}=I_{n}-\sum_{j=1}^{p-1}\Pi_{j}$.
Under such assumptions, it is well known that the elements of $Y_{t}$ are
individually, at most, $I(1)$ and that they are jointly cointegrated of order
$1$, in the sense that $\beta^{\prime}Y_{t-1}$\ is $I(0)$ (see
\cite{johansen1995likelihood} and the references therein).

To possibly reduce the number of parameters in the VECM, \cite{CM2024} take
the following assumptions:

\begin{assumption}
For $\Pi=[\Pi_{1}^{\prime},\ldots,\Pi_{p-1}^{\prime}]^{\prime}$ it holds%
\[
\Pi=A\omega^{\prime},
\]
where $\omega$ is a full-rank $n\times q$ matrix with $q<n$ and $A$ is a
full-rank $n(p-1)\times q$ matrix.
\end{assumption}

\begin{assumption}
It holds%
\[
\beta=\omega\gamma,
\]
where $\gamma$ is a full-rank $q\times r$ matrix with $q\geq r$.
\end{assumption}

Under Assumptions 4 and 5, Model (\ref{VECM}) can be rewritten in the
following {Vector-Error Correction Index Model [VECIM]:}%
\[
\Delta Y_{t}=\alpha_{0}\gamma^{\prime}f_{t-1}+\sum_{j=1}^{p-1}\alpha_{j}\Delta
f_{t-j}+\varepsilon_{t},
\]
where $\gamma$ is a full-rank $q\times r$ matrix ($q\geq r$), and $\alpha_{j}$
is an $n\times q$ matrix for $j=1,...,p-1$ such that $\mathrm{rank}%
([\alpha_{1}^{\prime},...,\alpha_{p-1}^{\prime}]^{\prime})=q$. Notice that the
cointegration matrix is given by $\beta=\omega\gamma$.

Interestingly, the indexes $f_{t}$ themselves are generated by a
$q-$dimensional VECM:%
\[
\Delta f_{t}=\underline{\alpha}_{0}\gamma^{\prime}f_{t-1}+\sum_{j=1}%
^{p-1}\underline{\alpha}_{j}\Delta f_{t-j}+\varepsilon_{t}^{\chi},
\]
where $\underline{\alpha}_{j}=\omega^{\prime}\alpha_{j}$, for $j=0,1\ldots
,p-1$.

By first inserting the decomposition (\ref{decomp}) between $\Psi(L)$ and
$\varepsilon_{t}$ into the Wold representation of the first differences
$\Delta Y_{t}$:%
\[
\Delta Y_{t}=\Psi(L)\varepsilon_{t},
\]
and then further decomposing the common component $\chi_{t}$ into permanent
and transitory subcomponents as in \cite{centoni2003measuring}, we get the
following:%
\begin{equation}
Y_{t}=\chi_{t}+\iota_{t}=\pi_{t}+\tau_{t}+\iota_{t}, \label{decomp_2}%
\end{equation}
where%
\begin{align}
\Delta\pi_{t}  &  =\Psi(L)\Sigma\omega\underline{\Sigma}^{-1}\underline{\Sigma
}\underline{\alpha}_{0\perp}(\underline{\alpha}_{0\perp}^{\prime
}\underline{\Sigma}\underline{\alpha}_{0\perp})^{-1}\underset{\varepsilon
_{t}^{\pi}}{\underbrace{\underline{\alpha}_{0\perp}^{\prime}\varepsilon
_{t}^{\chi}}},\label{Delta_pi}\\
\Delta\tau_{t}  &  =\Psi(L)\Sigma\omega\underline{\Sigma}^{-1}%
\underline{\alpha}_{0}(\underline{\alpha}_{0}^{\prime}\underline{\Sigma}%
^{-1}\underline{\alpha}_{0})^{-1}\underset{\varepsilon_{t}^{\tau
}}{\underbrace{\underline{\alpha}_{0}^{\prime}\underline{\Sigma}%
^{-1}\varepsilon_{t}^{\chi}}},\label{Delta_tao}\\
\Delta\iota_{t}  &  =\Psi(L)\omega_{\perp}(\omega_{\perp}^{\prime}\Sigma
^{-1}\omega_{\perp})^{-1}\underset{\varepsilon_{t}^{\iota}}{\underbrace{\omega
_{\perp}^{\prime}\Sigma^{-1}\varepsilon_{t}}} \label{Delta_iota}%
\end{align}

Since errors $\varepsilon_{t}^{\pi}$ are the innovations of the common trends
of the indexes $f_{t}$ (see e.g. \cite{johansen1995likelihood}) and errors
$\varepsilon_{t}^{\tau}$ are such that $\mathrm{E}(\varepsilon_{t}^{\pi
}\varepsilon_{t}^{\tau\prime})=0$, \cite{CM2024} label $\pi_{t}$\ as the
common permanent component, $\tau_{t}$\ as the common transitory component,
whereas $\iota_{t}$\ is the uncommon component given that $\mathrm{E}%
(\varepsilon_{t}^{\iota}\varepsilon_{t}^{\pi\prime})=0$ and $\mathrm{E}%
(\varepsilon_{t}^{\iota}\varepsilon_{t}^{\tau\prime})=0$.

Following a similar reasoning as the one leading to Proposition 1,
post-multiplying with $\omega_{\perp}$\ both sides of the relation%
\[
\Psi(L)(\Delta I_{n}-\sum_{j=1}^{p-1}\alpha_{j}\omega^{\prime}\Delta
L^{j}-\alpha_{0}\gamma^{\prime}\omega^{\prime}L)=\Delta I_{n}%
\]
we again get $\Psi(L)\omega_{\perp}=\omega_{\perp}$, which in turn implies
that the Wold polynomial matrix of the VECIM has the same form as
(\ref{Psi(L)}). Finally, inserting (\ref{Psi(L)}) in Equations (\ref{Delta_pi}%
), (\ref{Delta_tao}), and (\ref{Delta_iota}) we can prove the following proposition:

\begin{proposition}
In the VECIM, the first differences of the components of $Y_{t}$\ in
(\ref{decomp_2}) read%
\begin{align*}
\Delta\pi_{t}  &  =(\Sigma\omega\underline{\Sigma}^{-1}+\sum_{j=1}^{\infty
}\theta_{j}L^{j})\Sigma\underline{\alpha}_{0\perp}(\underline{\alpha}_{0\perp
}^{\prime}\underline{\Sigma}\underline{\alpha}_{0\perp})^{-1}\varepsilon
_{t}^{\pi}\equiv P(L)\varepsilon_{t}^{\pi},\\
\Delta\tau_{t}  &  =(\Sigma\omega\underline{\Sigma}^{-1}+\sum_{j=1}^{\infty
}\theta_{j}L^{j})\underline{\alpha}_{0}(\underline{\alpha}_{0}^{\prime
}\underline{\Sigma}^{-1}\underline{\alpha}_{0})^{-1}\varepsilon_{t}^{\tau
}\equiv T(L)\varepsilon_{t}^{\tau}\\
\Delta\iota_{t}  &  =\omega_{\perp}(\omega_{\perp}^{\prime}\Sigma^{-1}%
\omega_{\perp})^{-1}\varepsilon_{t}^{\iota},
\end{align*}
where the uncommon component $\iota_{t}$ is a $n-$dimensional random walk such
that $\mathrm{Rank}\left(  \mathrm{E}(\Delta\iota_{t}\Delta\iota_{t}^{\prime
})\right)  =n-q$.
\end{proposition}

Notice that Proposition 2 implies that Corollary 1 applies to the VECIM as well.

\begin{remark}
Given that the components in (\ref{decomp_2}) are not correlated with each
other at any lag and lead, the VECIM allows one to perform a structural
analysis taking advantage of the features of both the DFM, namely isolating
shocks that are common among variables, and the VECM, namely disentangling
shocks having transitory or permanent effects. For instance, one may identify
the structural transitory shocks as $u_{t}=C^{-1}D\varepsilon_{t}^{\tau}$ and
the impulse response functions as $\Theta(L)=T(L)D^{-1}C$, where $D$ is the
matrix formed by the first $r$ rows of $T(0)$ and $C$ is a lower triangular
matrix such that
\[
CC^{\prime}=D\underline{\alpha}_{0}^{\prime}\Sigma^{-1}\omega^{\prime}%
\Sigma\omega\Sigma^{-1}\underline{\alpha}_{0}D^{\prime}%
\]
Since the first $r$ rows of $\Theta(0)$, being equal to $C$, form a lower
triangular matrix, the usual interpretation of structural shocks obtained
through a Cholesky factorization applies to $u_{t}$.
\end{remark}

\cite{CM2024} offer a three-step SA for the estimation of the VECIM and
propose to select the triple $(p,q,r)$ in a unique search by IC. An extensive
Monte Carlo study shows that the proposed methodology works reasonably well
for $n$ ranging from $6$ to $18$ when the model is identified by the
Hannan-Quinn IC. Moreover, in an empirical application, they identify a shock
that maximizes the variability of the common transitory component of
unemployment at the business cycle frequencies, and another one that does the
same, but for the common permanent component of unemployment. These two shocks
are endowed with a neater economic interpretation than a unique main business
cycle shock identified according to \cite{angeletos2020business}.


\section{A New Proposal: The Cointegrated Index-Augmented Autoregressive
Model}

A possible limitation of the VECIM is that the uncommon component $\iota_{t}$
is necessarily a random walk, which may be considered restrictive for some
applications. For example, \cite{barigozzi2021large} propose a DFM where the
idiosyncratic components may be I(0) or I(1).

In order to overcome this issue, one can combine the VECIM with the IAAR.
Formally, this involves taking Assumption 5 along with the following one:

\begin{assumption}
For the VECM (\ref{VECM}) it holds%
\[
\pi_{ik}^{(j)}=%
{\textstyle\sum\limits_{m=1}^{q}}
\alpha_{im}^{(j)}\omega_{km},
\]
where $\pi_{ik}^{(j)}$ is the generic element of the polynomial matrix
$\Pi_{j}$ for $j=1,\ldots,p-1$, $i=1,\ldots,n$, $k=1,\ldots,i-1,i+1,\ldots,n$.
\end{assumption}

Taking Assumptions 5 and 6, the model (\ref{VECM}) can be rewritten into the
following Cointegrated Index-Augmented Auto-Regressive model [CIAAR]%
\begin{equation}
\Delta Y_{t}=\sum_{j=1}^{p-1}D_{j}\Delta Y_{t-j}+\alpha_{0}\gamma^{\prime
}\omega^{\prime}Y_{t-1}+\sum_{j=1}^{s-1}\alpha_{j}\omega^{\prime}\Delta
Y_{t-j}+\varepsilon_{t}, \label{CIAAR}%
\end{equation}
where $D_{j}$\ is a $n\times n$ diagonal matrix with $\delta_{ii}^{(j)}%
=\pi_{ii}^{(j)}-\sum_{m=1}^{q}\alpha_{im}^{(j)}\omega_{im}$\ as generic
diagonal element.

When the elements of the series $Y_{t}$ are I(1), Model (\ref{CIAAR}) includes
several earlier models as special cases. In fact, we have a MAI for series
$\Delta Y_{t}$ if $p,r=0$, an IAAR for series $\Delta Y_{t}$ is obtained if
$r=0$, and a VECIM if $p=0$.

\begin{remark}
Interestingly, by the same argument underlying Proposition 2, we see that,
differently from the VECIM, $\Psi(L)\omega_{\perp}\neq\omega_{\perp}$, which
implies, in view of Equation (\ref{Delta_iota}), that the first differences of
the uncommon component $\Delta\iota_{t}$\ are generally autocorrelated in the
case of the CIAAR. The uncommon component $\iota_{t}$ is still stochastically
singular with rank $n-q$. Since system (\ref{CIAAR}) has overall $n-r$ unit
roots, while common component $\chi_{t}$ has $q-r$ unit roots, uncommon
component $\iota_{t}$ has $n-q$ unit roots (see \cite{Deistler2017} and
\cite{barigozzi2020cointegration} on the properties of singular I(1)
stochastic processes).
\end{remark}

Following \cite{cubadda2019representation} and \cite{CM2024}, the estimation
procedure is based on an SA where each step is designed to increase the
Gaussian likelihood of Model (\ref{CIAAR}). In detail, when $0<r<q$, the
procedure goes as follows:

\begin{enumerate}
\item Given (initial) estimates of $\gamma$, $\omega$, and $D=[D_{1}%
,\ldots,D_{p-1}]^{\prime}$, maximize the conditional Gaussian likelihood $%
\mathcal{L}%
(A^{\dagger},{\Sigma}|\gamma,\omega,D)$ by estimating $A^{\dagger}=[\alpha
_{0}^{\prime},A^{\prime}]^{\prime}$, where $A=[\alpha_{1}^{\prime}%
,...,\alpha_{s-1}^{\prime}]^{\prime}$, and ${\Sigma}$ with OLS on the
following equation%
\[
\Delta Y_{t}-\sum_{j=1}^{p-1}D_{j}\Delta Y_{t-j}=\alpha_{0}\gamma^{\prime
}\omega^{\prime}Y_{t-1}+\sum_{j=1}^{s-1}\alpha_{j}\omega^{\prime}\Delta
Y_{t-j}+\varepsilon_{t}%
\]

\item Premultiply by ${\Sigma}^{-1/2}$ and apply the $\mathrm{Vec}$ operator
to both the sides of Equation (\ref{CIAAR}), then use the property
$\mathrm{Vec}(ABC)=(C^{\prime}\otimes A)\mathrm{Vec}(B)$ to get%
\[
{\Sigma}^{-1/2}\Delta Y_{t}=%
{\textstyle\sum\limits_{j=1}^{p-1}}
(Y_{t-j}^{\prime}\otimes\Sigma^{-1/2})\mathrm{Vec}(D_{j})+(Y_{t-1}^{\prime
}\otimes{\Sigma}^{-1/2}\alpha_{0}\gamma^{\prime}+%
{\textstyle\sum\limits_{j=1}^{s-1}}
\Delta Y_{t-j}^{\prime}\otimes{\Sigma}^{-1/2}\alpha_{j})\mathrm{Vec}%
(\omega^{\prime})+{\Sigma}^{-1/2}\varepsilon_{t},
\]
and reparametrize the above model as
\begin{equation}
{\Sigma}^{-1/2}\Delta Y_{t}=%
{\textstyle\sum\limits_{h=1}^{p-1}}
[(Y_{t-j}^{\prime}\otimes\Sigma^{-1/2})M]\delta_{j}+(Y_{t-1}^{\prime}%
\otimes{\Sigma}^{-1/2}\alpha_{0}\gamma^{\prime}+%
{\textstyle\sum\limits_{j=1}^{s-1}}
\Delta Y_{t-j}^{\prime}\otimes{\Sigma}^{-1/2}\alpha_{j})\mathrm{Vec}%
(\omega^{\prime})+{\Sigma}^{-1/2}\varepsilon_{t},\nonumber
\end{equation}
where $\delta_{j}$ is a $n-$vector such that $D_{j}=\mathrm{diag}(\delta_{j}%
)$, and $M$ is a binary $n^{2}\times n-$matrix whose generic element $m_{ik}$
is such that%
\[
m_{ik}=\left\{
\begin{array}
[c]{cc}%
1 & \text{if }i=1+(k-1)(n+1),\hspace{0.3cm}k=1,...,N\\
0 & \text{otherwise}%
\end{array}
\right.
\]
Given the previously obtained estimates of $A^{\dagger}$, $\gamma$, and
${\Sigma}$, maximize $%
\mathcal{L}%
(\omega,D|A^{\dagger},\gamma,{\Sigma})$ by estimating $\mathrm{Vec}%
(\omega^{\prime})$ and $\delta=[\delta_{1}^{\prime},...,\delta_{p-1}^{\prime
}]^{\prime}$ with OLS on Equation (\ref{Step_2}).

\item Given the previously obtained estimates of $\omega$ and $D$, maximize $%
\mathcal{L}%
(\gamma|\omega,D)$ by estimating $\gamma$ as the eigenvectors that correspond
to the $r$ largest eigenvalues of the matrix%
\[
S_{11}^{-1}S_{10}S_{00}^{-1}S_{01}%
\]
where $S_{ij}=\sum_{t=p+1}^{T}R_{i,t}R_{j,t}^{\prime}$\ for $i,j=0,1$,
$R_{0,t}$ and $R_{1,t}$ are, respectively, the residuals of an OLS regression
of $\Delta Y_{t}-\sum_{j=1}^{p-1}D_{j}\Delta Y_{t-j}$\ and $\omega^{\prime
}Y_{t-1}$ on $[\Delta Y_{t-1}^{\prime}\omega,\ldots,\Delta Y_{t-s+1}^{\prime
}\omega]^{\prime}$.

\item Repeat steps 1 to 3 until numerical convergence occurs.
\end{enumerate}

When $r=0$, step 3 is clearly not needed, and steps 1 and 2 must be modified
as follows:

\begin{enumerate}
\item[1.1] Given (initial) estimates of $\omega$ and $D$, maximize $%
\mathcal{L}%
(A,{\Sigma}|\omega,D)$ by estimating $A$ and ${\Sigma}$ with OLS on the
following model%
\[
\Delta Y_{t}-\sum_{j=1}^{p-1}D_{j}\Delta Y_{t-j}=\sum_{j=1}^{s-1}\alpha
_{j}\omega^{\prime}\Delta Y_{t-j}+\varepsilon_{t}%
\]

\item[2.1] Given the previously obtained estimates of $A$ and ${\Sigma}$,
maximize $%
\mathcal{L}%
(\omega|A,{\Sigma})$ by estimating $\mathrm{Vec}(\omega^{\prime})$ and
$\delta$\ with OLS on the following model%
\[
{\Sigma}^{-1/2}\Delta Y_{t}=%
{\textstyle\sum\limits_{h=1}^{p-1}}
[(Y_{t-j}^{\prime}\otimes\Sigma^{-1/2})M]\delta_{j}+(%
{\textstyle\sum\limits_{j=1}^{s-1}}
\Delta Y_{t-j}^{\prime}\otimes{\Sigma}^{-1/2}\alpha_{j})\mathrm{Vec}%
(\omega^{\prime})+{\Sigma}^{-1/2}\varepsilon_{t},
\]

\end{enumerate}

Finally, when $r=q$, we can assume without loss of generality that
$\gamma=I_{q}$. Then Step 3 is again not needed, whereas Steps 1 and 2 must be
modified as follows.

\begin{enumerate}
\item[1.3] Given (initial) estimates of $\omega$ and $D$, maximize $%
\mathcal{L}%
(A^{\dagger},{\Sigma}|\omega,D)$ by estimating $A^{\dagger}$ and ${\Sigma}$
with OLS on the following model%
\[
\Delta Y_{t}-\sum_{j=1}^{p-1}D_{j}\Delta Y_{t-j}=\alpha_{0}\omega^{\prime
}Y_{t-1}+\sum_{j=1}^{s-1}\alpha_{j}\omega^{\prime}\Delta Y_{t-j}%
+\varepsilon_{t}%
\]

\item[2.3] Given the previously obtained estimates of $A^{\dagger}$ and
${\Sigma}$, maximize $%
\mathcal{L}%
(\omega,D|A^{\dagger},{\Sigma})$ by estimating $\mathrm{Vec}(\omega^{\prime})$
and $\delta$\ with OLS on the following model%
\[
{\Sigma}^{-1/2}\Delta Y_{t}=%
{\textstyle\sum\limits_{h=1}^{p-1}}
[(Y_{t-j}^{\prime}\otimes\Sigma^{-1/2})M]\delta_{j}+(Y_{t-1}^{\prime}%
\otimes{\Sigma}^{-1/2}\alpha_{0}+%
{\textstyle\sum\limits_{j=1}^{s-1}}
\Delta Y_{t-j}^{\prime}\otimes{\Sigma}^{-1/2}\alpha_{j})\mathrm{Vec}%
(\omega^{\prime})+{\Sigma}^{-1/2}\varepsilon_{t}%
\]

\end{enumerate}

The choice of initial values for the above procedures is discussed in the
Appendix, whereas the selection of the quadruple $(p,s,q,r)$ can be done by
IC, sequentially or in a unique search as suggested by \cite{CM2024}.


\section{Conclusions}

The DFM and the VAR are, arguably, among the most popular tools in
macroeconometrics and financial econometrics. The two approaches should be
considered complementary rather than substitutive, since each of them has its
own merits. The MAI represents a link between these two methodologies: On the
one hand, it is a VAR with a specific reduced-rank structure that alleviates
the dimensionality problem; on the other hand, the MAI and its variants have
several analogies with the DFM; in particular, they allow for identifying a
small number of common reduced-form errors and for recovering structural
shocks from those errors only.

However, the MAI is not affected by some theoretical limitations of the DFM
such as the requirement that the cross-sectional dimension diverges to
infinity and the need for specific assumptions on the dynamic correlation
structure of the idiosyncratic component and on the factor loadings. In a more
practical perspective, VARs with an index structure can also handle features
such as stochastic volatility (\cite{carriero2022global}) and time-varying
parameters (\cite{CGG2025}), which are not easily accommodated in DFMs.

Recent developments in VAR models with index structures have considerably
extended the original MAI formulation, endowing the model with individual
autoregressive structures, stochastic volatility, time-varying parameters,
high dimensionality, and cointegration. These extensions have proven to be
useful tools for detecting common components, obtaining efficiency gains
through the imposition of parameter restrictions, performing structural
analysis, and boosting forecast accuracy.

Having reviewed most of the recent advances on the MAI and provided new
insights on the representation theory underlying the various formulations, a
new model, namely the CIAAR, has been proposed along with an estimation
procedure. The hope is that this paper will contribute to providing room for
future research on VAR models with index structures.


\section{Appendix}

The choice of the initial values for a SA is important. Not only is an
accurate initialization necessary to boost numerical convergence, but the SA
is asymptotically equivalent to the ML one when the parameters to be
initialized are consistently estimated (\cite{HautschJoFE}).

With reference to the SA in Section 3, the initial values for $\gamma$,
$\omega$, and $D$ can be obtained as follows:

\begin{enumerate}
\item Use the usual Johansen procedure on the model (\ref{VECM}) and obtain
estimates $\hat{\alpha}_{0}$, $\hat{\beta}$, and $\hat{\Pi}_{j}$ for
$j=1,\ldots,m$, where $m=\max\{p,s\}-1$.

\item Construct matrices $\tilde{\Pi}_{j}=\hat{\Pi}_{j}-\mathrm{diag}[\hat
{\pi}_{11}^{(j)},\ldots,\hat{\pi}_{nn}^{(j)}]^{\prime}$ for $j=1,\ldots,m$.

\item Construct the matrix $\tilde{\Phi}=[\tilde{\Pi}_{1}^{\prime}%
,\ldots,\tilde{\Pi}_{m}^{\prime},\hat{\beta}\hat{\alpha}_{0}^{\prime
},]^{\prime}$.

\item Compute the singular value decomposition $\tilde{\Phi}=U\Lambda
V^{\prime}$, where the singular values are not increasingly ordered, and
obtain $\hat{\omega}$ as the matrix formed by the first $q$ columns of $V$.

\item Compute the $q-$rank approximation of $\tilde{\Phi}$ as $\bar{\Phi
}=U\bar{\Lambda}V^{\prime}$, where $\bar{\Lambda}$ is obtained from $\Lambda$
by setting to $0$ the smallest singular values of $n-q$.

\item Construct $\bar{\Pi}=[\bar{\Pi}_{1}^{\prime},\ldots,\bar{\Pi}%
_{p-1}^{\prime}]^{\prime}$ as the matrix formed by the first $n(p-1)$ rows of
$\bar{\Phi}$.

\item Construct $\hat{D}_{j}$ as a diagonal matrix with the diagonal equal to
$\mathrm{diag}[\hat{\pi}_{11}^{(j)}-\bar{\pi}_{11}^{(j)},\ldots,\hat{\pi}%
_{nn}^{(j)}-\bar{\pi}_{nn}^{(j)}]^{\prime}$ for $j=1,\ldots,s-1$.
\end{enumerate}

The motivation for the above choices is twofold. First, the asymptotic
distribution of the Johansen estimator of $\beta$ is not affected by
restrictions on the short-run parameters (\cite{johansen1995likelihood}),
which implies that $\hat{\alpha}_{0}$, $\hat{\Pi}_{j}$ and $\tilde{\Pi}_{j}$
are consistent, although inefficient, estimators of the associated parameters.
Second, the right-singular vectors that correspond to the $q$ largest singular
values of the matrix $\tilde{\Phi}$ consistently estimate $\omega$ (see, e.g.,
\cite{Reinsel2022book}). By the same argument, $\bar{\Pi}$ provides a
consistent estimator of $\Pi$. Finally, the consistency of $\hat{D}=[\hat
{D}_{1},\ldots,\hat{D}_{p-1}]^{\prime}$\ trivially follows from the ones of
$\hat{\Pi}$\ and $\bar{\Pi}$.

\bibliographystyle{apalike}
\bibliography{biblio}

\end{document}